\documentstyle[12pt]{article}
\newcommand{\del}{\partial}
\title{ \vspace*{-10mm}
\begin{tabular}{ll}
\hspace*{8cm} & {\normalsize LANCS-TH} \vspace{-3mm} \\
\hspace*{8cm} & {\normalsize May 2000}
\end{tabular} \vspace{18mm} \\
\bf Gravity localization with a domain wall junction in six dimensions} 
\author{Takeshi Nihei\\
 \mbox{} \vspace{3mm}\\
 \normalsize \em  Department of Physics, Lancaster University, Lancaster LA1 4YB, UK\\
\verb+t.nihei@lancaster.ac.uk+ }
\date{ }
\begin{document}
\baselineskip 6mm
\renewcommand{\thefootnote}{\fnsymbol{footnote}}
\begin{titlepage}
\maketitle
\thispagestyle{empty}
\begin{abstract}
\normalsize
\baselineskip 6mm
We study gravity localization in the context of a six-dimensional 
gravity model coupled with complex scalar fields.
With a supergravity-motivated scalar potential, 
we show that the domain wall junction solutions localize 
a four-dimensional massless graviton under an assumption on 
the wall profile. 
We find that unlike the global supersymmetric model, 
contributions to the junction tension cancel locally with 
gravitational contributions. 
The wall tension vanishes due to the metric suppression. 
\end{abstract}
\end{titlepage}
\newpage
%
%
\section{Introduction}
%
%

Possibilities of extra dimensions have been studied for a long time,
and the existence is well motivated by superstring theories. 
A traditional method to hide extra dimensions is an idea of 
compactification in which the extra dimensions are supposed to be 
extremely small.

In a couple of years, it has been recognized that 
the extra dimensions may have sub-millimeter size \cite{large-extra-dim} or
infinitely large volume \cite{Randall-Sundrum-1,Randall-Sundrum-2,Lykken-Randall} if the standard model fields are confined at a three brane.
In particular Randall and Sundrum recently proposed 
an scenario alternative to compactification using exponentially 
warped metric in a five-dimensional gravity 
model \cite{Randall-Sundrum-1,Randall-Sundrum-2}. 
Due to the warped metric, the four-dimensional massless graviton
are localized on a brane, and the four-dimensional Newton law 
is approximately realized on the brane at low energy. 
Supersymmetrization of this scenario has been discussed in 
Ref. \cite{Bagger-etal}.

In the Randall-Sundrum scenario, 
cosmological constants are introduced for both 
the five-dimensional bulk and the four-dimensional branes, and 
the warped metric is derived as a solution to the Einstein 
equation \cite{Visser+Squires,Gogbera}.
In order to obtain the solution with the four-dimensional
Poincar\'{e} invariance, however, 
the cosmological constants have to be specially tuned. 
Stabilization mechanism of the extra dimension has been discussed
by introducing a bulk scalar field \cite{Goldberger-Wise}.

In order to have a natural explanation of this tuning, 
we have to discuss the origin of the brane cosmological constant. 
Instead of pursuing string theories, we consider a 
the field theoretic approach in this paper. 
Several works using domain wall solutions in five-dimensional gravity
models have been done 
\cite{Behrndt-Cvetic,DeWolfe-etal,Gremm,Verlinde}. 
In these analyses, a supergravity-motivated scalar potential
is introduced \cite{5D-N2-gauged-SUGRA}, and the domain wall 
solution of the scalar fields
produces an effective cosmological constant on the brane 
to implement the warped metric in the Randall-Sundrum scenario. 
This line is a gravity version of a previous idea of living in 
a domain wall \cite{Akama,Ruba-Shapo,Dva-Shif}. 
An analysis on domain walls in arbitrary dimensions is 
given in Ref. \cite{Skenderis-Townsend}. 
Domain wall solutions in four-dimensional supergravity
models have been studied in Ref. \cite{4D-SUGRA-domain-wall}.

However it has recently been pointed out that smooth domain wall 
solutions interpolating between supersymmetric vacua can not exist
in odd-dimensional supergravity models \cite{Odd-dimension}. 
This naturally leads us to work in a framework of six-dimensional 
supergravity models \cite{6D-SUGRA}. 
In this case we need a two-dimensional topological solution
like a vortex solution \cite{Nielsen-Olesen}. 
Gravity localization on a string-like defect in six dimensions
has been studied in Ref. \cite{string-like-defect}.

There is another interesting two-dimensional stable solitonic solution, 
namely, a domain wall junction solution in supersymmetric 
models \cite{Abra-Town,Gibb-Town,quarter-BPS,Gors-Shif,exact-junction-solution,Bino-Veld}. 
The domain wall junction preserves one-quarter of the underlying 
supersymmetry \cite{Gibb-Town,quarter-BPS}, and satisfies
a first order BPS equation \cite{BPS}.
With domain wall junctions, general analyses on gravity localization 
in infinitely large extra dimensions have been given
in Refs. \cite{co-dimension-one,Junction,Csaki-etal}.

In this paper we study gravity localization in the context of
a six-dimensional gravity model coupled with complex scalar 
fields. Similar analyses have been done 
in Refs. \cite{Carroll-etal,Network}.
With a supergravity-motivated scalar potential, 
we derive first order equations which the metric and the scalar 
fields should satisfy. We calculate the tensions of domain wall
and juncion. Finally we study a warped metric and discuss gravity 
localization on the junction.

\vspace{5mm}
%
%
%
\section{Set-up}
%
%

We consider a six-dimensional gravity coupled with 
complex scalar fields $\phi^i$ ($i$ $=$ 1, $\cdots$, $N$). 
We use coordinates $x^M$ $=$ $(x^\mu, x^m)$, 
where $M$ $=$ $0,1,2,3,5,6$, $\mu$ $=$ $0,1,2,3$ and $m$ $=$ $5,6$. 
$x^\mu$ are coordinates for
the observable four spacetime dimensions. $x^m$ are coordinates for 
the two extra dimensions, where $-\infty$ $<$ $x^m$ $<$ $\infty$. 
The action is given by 
\begin{eqnarray}
S & = & \int d^6 x
        \sqrt{-g} \left[ -\frac{1}{2 \kappa^2} R 
        + K_{ij^*}g^{MN}\del_M \phi^i \del_N \phi^{j*}
        - V(\phi,\phi^*) \right], 
\label{eqn:action}
\end{eqnarray}
where $g_{MN}$ is the metric in a time-like 
signature convention ($+$, $-$, $-$, $-$, $-$, $-$). 
$V(\phi,\phi^*)$ is a scalar potential. 
The scalar kinetic term has field dependent coefficients
$K_{ij^*}(\phi,\phi^*)$ which are derived from the K\"{a}hler
potential as $K_{ij^*}$ $=$ 
$\del^2 K(\phi,\phi^*)/\del \phi^i \del \phi^{j*}$. 
In the following, we adopt the six-dimensional Planck mass unit
$\kappa^2$ $=$ 1 unless otherwise stated. 
We put the following ansatz for the background metric
\begin{eqnarray}
ds^2 & = & g_{MN} dx^M dx^N \nonumber \\
     & = & e^{2 \sigma(x^5, \,x^6)} \left[ \eta_{\mu \nu} dx^{\mu} dx^{\nu} 
                            - (dx^5)^2 -  (dx^6)^2 \right],
\label{eqn:metric-ansatz}
\end{eqnarray}
where $\eta_{\mu \nu}$ is a four-dimensional flat metric. 
This metric guarantees the four-dimensional Poincar\'{e} invariance. 
We look for static scalar configurations which depend 
on $x^m$ only: $\phi^i$ $=$ $\phi^i(x^5, x^6)$. 
Then the equations of motion can be written as
\begin{eqnarray}
\del_m^2 \sigma + 4 (\del_m \sigma)^2 &  = & - \frac{1}{2} e^{2 \sigma} V, 
\label{eqn:eom1} \\
-4 \del_m \del_n \sigma + 4 (\del_m \sigma)(\del_n \sigma) & = & 
K_{ij^*} (\del_m \phi^i \del_n \phi^{j*}
              +\del_n \phi^i \del_m \phi^{j*}), 
\label{eqn:eom2} \\
\del_m^2 \phi^i + 4 (\del_m \sigma) (\del_m \phi^i) 
        \hspace{2.2cm} \nonumber \\
+ \, K^{ij^*} (\del_k K_{lj^*})(\del_m \phi^k)(\del_m \phi^l)
 & = & e^{2 \sigma} K^{ij^*} \frac{\del V}{\del \phi^{j*}},
\label{eqn:eom3}
\end{eqnarray}
where repeated indices are summed. 
The first two equations correspond to $(\mu,\nu)$ and $(m,n)$
components of the Einstein equations, respectively. The last one
is the scalar field equation.

For the scalar potential, we assume the following form motivated by 
supergravity models \cite{Behrndt-Cvetic,Skenderis-Townsend,positive-energy-theorem}:
\begin{eqnarray}
V & = & e^{K} \left[ K^{ij^*} (D_i W)(D_j W)^*
- \frac{5}{2}|W|^2 \right], 
\label{eqn:potential}
\end{eqnarray}
where $D_i W$ $=$ $\del W/\del \phi^i$ $+$ $(\del K/\del \phi^i)W$ 
and $W$ is an arbitrary function of $\phi^i$ which may be interpreted
as a superpotential if we can derive this potential from a 
supergravity model. 
A five-dimensional gauged supergravity model has a similar 
potential \cite{5D-N2-gauged-SUGRA}. 
Given the potential (\ref{eqn:potential}), we can obtain the first 
order equations which the metric and the scalar fields should 
satisfy \cite{DeWolfe-etal,Behrndt-Cvetic,Verlinde,Skenderis-Townsend,Carroll-etal,Network}. 
We will see this in the next section. 
These first order equations make it easy to solve the equations of
motion (\ref{eqn:eom1})-(\ref{eqn:eom3}).

In general the potential (\ref{eqn:potential}) is not bounded below
because of $|W|^2$ term. Furthermore in some higher-dimensional
supergravity models, there are no local minima and 
all the extrema are local maxima or saddle points. 
However it has been shown that even a vacuum at a maximum is stable under 
local fluctuations around an AdS background unless the curvature of the
potential at the maximum is too negative \cite{positive-energy-theorem}. 
In this sense the potential (\ref{eqn:potential}) is sensible.

No six-dimensional supergravity models which provide the potential 
(\ref{eqn:potential}) have been constructed, so we can not 
work in a supergravity context. In this paper we just assume 
the potential (\ref{eqn:potential}) from the beginning.

\vspace{5mm}
%
%
%
\section{Tensions of domain wall and junction}
%
%

In this section, we derive the first order equations for the metric
and the scalar fields. We also obtain the formula for the tensions
of domain wall and junction.

With the potential (\ref{eqn:potential}), we can write the action 
as a sum of perfect squares up to total
derivative terms \cite{BPS}. For this purpose, 
it is convenient to define a complex coordinate variable 
$z$ $=$ $x^5$ $+$ $i x^6$ and derivatives $\del$ $=$  
$(\del_5 - i \del_6)/2$, $\bar{\del}$ $=$ $(\del_5 + i \del_6)/2$. 
In terms of this variable, the action (\ref{eqn:action}) is written as
\begin{eqnarray}
S & = & \int d^6 x \, e^{6 \sigma} \left[ - 20 e^{-2 \sigma}
\left( \bar{\del} \del \sigma + 2 \del \sigma \bar{\del}\sigma \right)
-2 e^{-2 \sigma} K_{ij^*} \left( \del \phi^i \bar{\del} \phi^{j*}
+ \bar{\del}\phi^i \del \phi^{j*} \right) \right. \nonumber \\
 &   & \hspace{6cm} \left.
- e^{K} \left\{ K^{ij^*} (D_i W)(D_j W)^* 
- {\textstyle \frac{5}{2}}|W|^2 \right\} \right].
\label{eqn:action2}
\end{eqnarray}
We can make a perfect square, e.g., from
a part of the $\del \sigma \bar{\del}\sigma$ term 
and the $|W|^2$ term. In doing that we 
can introduce a complex phase $\theta(x^5,x^6)$ as follows:
\begin{eqnarray}
\left| \del \sigma - {\textstyle \frac{1}{4}} 
e^{\sigma} e^{K/2} e^{i \theta} W^* \right|^2.
\label{eqn:squaring}
\end{eqnarray}
Taking into account this phase, the action can be written 
as a sum of a local contribution and a topological term
\cite{Skenderis-Townsend,BPS,Carroll-etal}
\begin{eqnarray}
S & = & S_{\rm local} + S_{\rm topological}.
\label{eqn:energy}
\end{eqnarray}
The local contribution is given by a sum of perfect squares
\begin{eqnarray}
S_{\rm local} & = & \int d^6 x \, e^{4 \sigma} \left[ 
40 |D \sigma|^2 - 4 K_{ij^*} (D \phi^i)(D \phi^j)^*
- \left\{ 4i (D \theta)(D \sigma)^* + {\rm h.c.} \right\} \right]
\nonumber \\
 & = &  \int d^6 x \, e^{4 \sigma} \left[ 40 \left|D \sigma 
-{\textstyle \frac{i}{10}} D \theta \right|^2 
- 4 K_{ij^*} (D \phi^i)(D \phi^j)^* 
- {\textstyle \frac{2}{5}} \left|D \theta \right|^2 \right],
\label{eqn:S_local}
\end{eqnarray}
where the symbols $D \sigma$, $D \phi^i$ and $D \theta$ are defined by
\begin{eqnarray}
D \sigma & = & \del \sigma 
     - {\textstyle \frac{1}{4}} e^{\sigma} e^{K/2} e^{i \theta} W^*,
\label{eqn:Dbps1} \\
D \phi^i & = & \del \phi^i 
     + {\textstyle \frac{1}{2}} K^{ij^*} 
e^{\sigma} e^{K/2} e^{i \theta} (D_j W)^*,
\label{eqn:Dbps2} \\
D \theta & = & \del \theta
               - {\textstyle \frac{i}{2}} \left( K_i \del \phi^i 
                                   - K_{i^*} \del \phi^{i*} \right).
\label{eqn:Dbps3}
\end{eqnarray}
The perfect squares in the integrand imply that 
the configurations which extremize the action satisfy the
first order equations $D \sigma$ $=$ $D \phi^i$ $=$ $D \theta$ $=$ 0,
namely, 
\begin{eqnarray}
\del \sigma & = & \frac{\kappa^2}{4} e^{\sigma} 
e^{\kappa^2 K/2} e^{i \theta} W^*, 
\label{eqn:BPS1} \\
\del \phi^i & = & - \frac{1}{2} K^{ij^*} e^{\sigma} 
e^{\kappa^2 K/2} e^{i \theta}
 (D_j W)^*,
\label{eqn:BPS2} \\
\del \theta & = & \frac{i}{2 \kappa^2}\left( K_i \del \phi^i 
                  - K_{i^*} \del \phi^{i*} \right).
\label{eqn:BPS3}
\end{eqnarray}
Here we have recovered the gravitational coupling explicitly. 
Note that the solutions to equations (\ref{eqn:BPS1})-(\ref{eqn:BPS3}) 
automatically satisfy the equations of 
motion (\ref{eqn:eom1})-(\ref{eqn:eom3}). 
In the four-dimensional supergravity model, similar equations are derived
from conditions for the existence of unbroken supersymmetry. 
The equation (\ref{eqn:BPS2}) implies that the scalar fields $\phi^i$ 
should depend on both $z$ and $z^*$ in order to have a non-trivial 
configuration. 

Because of the positive contribution $40 |D \sigma|^2$ in 
eq. (\ref{eqn:S_local}), the solutions to these first order equations
do not minimize the energy in general. However on an AdS background, 
the vacuum at an extremum is stable under local 
fluctuations unless the curvature of the
potential at the maximum is too negative \cite{positive-energy-theorem}.

The equation (\ref{eqn:BPS3}) can be written as $\del_m \theta$ $=$
$-{\rm Im}(K_i \del_m \phi^i)$. In the four dimensional supergravity model,
the quantity ${\rm Im}(K_i \del_m \phi^i)$ corresponds to an auxiliary 
vector field in a supergravity multiplet. This field acts as a 
gauge field for the K\"{a}hler-Weyl invariance.

The topological term consists of two kinds of total derivative terms
\begin{eqnarray}
S_{\rm topological} & = & - \int d^4 x \, \sum_m ( Z_m +  Y_m ).
\label{eqn:topological}
\end{eqnarray}
The first contributions $Z_m$ are given by
\begin{eqnarray}
Z_5 & = & \int dx^5 dx^6 \, \del_5 \left[ e^{4 \sigma} (5 \del_5 \sigma) 
  - e^{5 \sigma}e^{K/2}(e^{-i \theta}W + e^{i \theta}W^* ) \right], 
       \nonumber \\
Z_6 & = & \int dx^5 dx^6 \, \del_6 \left[ e^{4 \sigma} (5 \del_6 \sigma) 
  +i e^{5 \sigma}e^{K/2}(e^{-i \theta}W - e^{i \theta}W^* )\right].
\label{eqn:Z_m}
\end{eqnarray}
These include the domain wall tensions and the metric terms. 
The second contributions are
\begin{eqnarray}
Y_5 & = & - \int dx^5 dx^6 \, \del_5 \left[ e^{4 \sigma} 
\left\{ \del_6 \theta 
        + {\rm Im}(K_i \del_6 \phi^i) \right\} \right], \nonumber \\
Y_6 & = &  \int dx^5 dx^6 \, \del_6 \left[ e^{4 \sigma} 
\left\{ \del_5 \theta 
        + {\rm Im}(K_i \del_5 \phi^i) \right\} \right].
\label{eqn:Y_m}
\end{eqnarray}
These include the ordinary domain wall junction 
tentions ${\rm Im}(K_i \del_m \phi^i)$ and 
`gravitational' contributions $\del_m \theta$. 
Note that $Y_m$ can have non-vanishing values only when the fields
have two-dimensional non-trivial configurations, while $Z_m$ can be
non-vanishing even when the fields depends on only one of $x^m$.

In the absence of gravity, the derivatives of $\theta$ do not 
contribute, and the terms ${\rm Im}(K_i \del_m \phi^i)$ give 
the domain wall junction tension. 
Note that $\del_m \theta$ terms in eq. (\ref{eqn:Y_m}) are proportional
to $\kappa^2$ if we write the gravitational coupling explicitly. 
In the presence of gravity, however, the derivatives of $\theta$
contribute to $Y_m$. 
Substituting the first order equation (\ref{eqn:BPS3}) into 
eq. (\ref{eqn:Y_m}), we find that
the two contributions $\del_m \theta$ and ${\rm Im}(K_i \del_m \phi^i)$
cancel locally in the integrand with each other. 
Therefore the junction tentions $Y_m$ vanish
\begin{eqnarray}
Y_m & = & 0.
\label{eqn:Ym_eq_0}
\end{eqnarray}
This implies that in the presence of the gravitational degrees of 
freedom, a constant $\theta$ configuration can not extremize 
the action. The system requires a non-trivial $\theta$ to extremize
the action in such a way that $\del_m \theta$ cancel the ordinary
contributions ${\rm Im}(K_i \del_m \phi^i)$ to the junction tension. 

Let's explain briefly the reason why $\del_m \theta$ and 
${\rm Im}(K_i \del_m \phi^i)$ appear in the vanishing combination 
$D \theta$. 
In order to see this, it is essential to observe that the total 
derivative in the domain wall tension term produce 
the vanishing combination as follows:
\begin{eqnarray}
\del_m \left[ e^{K/2} e^{-i \theta} W \right] & = & 
e^{K/2} e^{-i \theta} \left[ (D_i W)\del_m \phi^i 
- i W (\del_m \theta + {\rm Im} K_i \del_m \phi^i ) \right].
\label{eqn:combination}
\end{eqnarray}
We can see that in the other parts of calculations, 
$\del_m \theta$ and ${\rm Im}(K_i \del_m \phi^i)$ always appear
in the same combination. 
If we work in the four-dimensional supergravity model, 
we may have deeper understanding of this result referring to
the K\"{a}hler-Weyl invariance.

As for the wall contributions $Z_m$, the situation is different. 
Substituting the first order equation (\ref{eqn:BPS1}) into 
eq. (\ref{eqn:Z_m}), we see that $Z_m$ are written as 
\begin{eqnarray}
Z_5 & = & \int dx^5 dx^6 \, \del_5 \left[ {\textstyle \frac{1}{2}}
e^{5 \sigma}e^{K/2} {\rm Re}(e^{-i \theta}W) \right], 
       \nonumber \\
Z_6 & = & \int dx^5 dx^6 \, \del_6 \left[ {\textstyle \frac{1}{2}}
  e^{5 \sigma}e^{K/2} {\rm Im}(e^{-i \theta}W) \right].
\label{eqn:Zm_2}
\end{eqnarray}
Unlike the junction tensions, the wall tensions $Z_m$ do not 
vanish in general.

\vspace{5mm}
%
%
%
\section{Warped metric from domain wall junction}
%
%

In this section, we discuss the solution to the first order 
equations (\ref{eqn:BPS1})-(\ref{eqn:BPS3}). 
In most cases the analytic solution is not available, so in this papar
we only discuss rough behavior of the solution. 
As an example for the function $W$, let's consider a quartic form 
for a single complex scalar field $\phi$
\begin{eqnarray}
W & = & -\frac{1}{4}\phi^4 + \phi. 
\label{eqn:superpot}
\end{eqnarray}
For the K\"{a}hler potential $K$, we take the minimal form 
$K$ $=$ $\phi \phi^*$. 
It is known that in the global supersymmetric model
such a quartic superpotential $W$ allows static domain wall junction 
solutions \cite{Abra-Town,Gibb-Town,quarter-BPS,Gors-Shif}. 
In the global case, the scalar potential has three
isolated degenerate minima at $\phi$ $=$ 1, $\omega$, $\omega^2$, 
where $\omega$ $=$ $e^{2 \pi i /3}$. 
Note that the potential does not allow static domain wall solutions 
in the four-dimensional supergravity model \cite{4D-SUGRA-domain-wall}.
This comes from the reality of the metric. In the domain wall junction 
case, however, the same discussion cannot be applied since the first
order equation (\ref{eqn:BPS1}) for the metric includes the derivative 
with respect to the complex coordinate $z$. 

With the function $W$ in eq. (\ref{eqn:superpot}) and 
the minimal K\"{a}hler potential $K$, we find three vacua 
$\phi$ $=$ $k$, $k \omega$, $k \omega^2$ by solving $D_{\phi} W$ $=$ 0. 
Here $k$ $\approx$ 1.2 is a single solution of a quintic equation
$k^5$ $+$ $4k^3$ $-4k^2$ $-4$ $=$ 0. 
The potential (\ref{eqn:potential}) and the first order equations
(\ref{eqn:BPS1})-(\ref{eqn:BPS3}) in this case are invariant 
under ${\bf Z}_3$ action $z$ $\rightarrow$ $\omega z$, 
$\phi$ $\rightarrow$ $\omega \phi$, $\sigma$ $\rightarrow$ $\sigma$,
$\theta$ $\rightarrow$ $\theta$. Therefore we expect a ${\bf Z}_3$ 
invariant domain wall junction solution. 
The solution describes the two-dimensional space separated into 
three regions by three walls, and these regions are labeled 
by the vacua $k$, $k \omega$, $k \omega^2$. 
Equations (\ref{eqn:BPS2}) and (\ref{eqn:BPS3}) imply that 
outside the walls and the junction the scalar fields $\phi$ and
the phase $\theta$ are almost constant. 
We consider the solution with arg$\phi$ $=$ arg$(-z)$ outside the wall
so that the junction is centered at $z$ $=$ 0. 
Then the first order equation (\ref{eqn:BPS1}) leads to the 
asymptotic behavior of the metric at spacial infinity in the extra 
dimensions
\begin{eqnarray}
e^{2\sigma} & \longrightarrow & \frac{C^2}{[{\rm Re}(\omega^{*n}z)]^2},
 \hspace{1cm} |z| \longrightarrow \infty,
\label{eqn:metric_AdS}
\end{eqnarray}
where $n$ $=$ 0, 1, 2 label the vacua $k$, $k \omega$, $k \omega^2$,
respectively. The constant $C$ is given by 
$C$ $=$ $2 e^{-k^2}(k-k^4/4)^{-1}$, where we have chosen $\theta$ $=$ 0 
outside the wall. 
The above metric (\ref{eqn:metric_AdS}) describes an AdS background. 
This asymptotic form is consistent with analyses in 
Refs. \cite{co-dimension-one,Junction,Carroll-etal}. 
From this behavior it follows that 
the extra dimension is infinitely large, since the volume 
$V$ $=$ $\int dx^5 dx^6$ $e^{2 \sigma}$ $\sim$ $\int dr/r$ is 
divergent where $r$ $=$ $|z|$.

Note that this behavior (\ref{eqn:metric_AdS}) can be applied 
only outside the walls.
Inside the wall the scalar field is not constant any more, 
hence we have to solve the coupled equations 
(\ref{eqn:BPS1})-(\ref{eqn:BPS3}) to know the profile. 
However it seems natural to assume that for a fixed $r$, 
the wall essentially has a kink-like profile. 
Then the evolution equation (\ref{eqn:BPS2}) along the direction 
perpendicular to the wall is given by $\del \phi$ $\sim$ $r^{-1}$
$(D_{\phi} W)^*$ inside the wall. 
This is consistent with the metric behavior 
(\ref{eqn:metric_AdS}) outside the wall. 
The metric may have a peak structure in the 
wall \cite{4D-SUGRA-domain-wall}, 
but the height of the peak scales as $1/r^2$. 
Thus the scaling property $e^{2\sigma}$ $\sim$ $1/r^2$ holds 
even in the walls. 
Also this assumption means that the wall width grows up 
like $\sim$ $r$ far from the origin. 
On the other hand the `height' of the wall remains the same 
even far from the origin. 
Consequently, the wall structure almost disappears far from 
the junction due to the metric suppression.

It is difficult to discuss the solution inside the junction, too.
However we can see the behavior of the metric near 
the origin $x^m$ $=$ 0 in a weak gravitational coupling approximation.
In the $\kappa^2$ $=$ 0 limit, the metric $\sigma$ and the phase $\theta$ 
are constant, and eq. (\ref{eqn:BPS2}) reduces to a simpler equation. 
From symmetry consideration the scalar field must vanish at the origin. 
Then it follows that $\phi$ $\approx$ $-z$ from eq. (\ref{eqn:BPS2}). 
In the first order in $\kappa^2$, the metric equation (\ref{eqn:BPS1})
becomes $\del \sigma$ $=$ $W^*/4$ where we have substituted
the zeroth order result in the right-hand side. 
Using this equation, we obtain $\sigma$ $\approx$ $-z z^*/4$ 
near the origin so that
\begin{eqnarray}
e^{2\sigma} & \approx & 1-\frac{z z^*}{2}, \hspace{1cm} |z| \ll 1.
\label{eqn:metric_near_origin}
\end{eqnarray}
The curvature near the origin is constant $R$ $\approx$ $-10$
up to the order of $|z|^2$.

Let's estimate the topological charges $Z_m$ in this example. 
Unlike the case of $Y_m$, the integrand of $Z_m$ in eq. (\ref{eqn:Zm_2})
does not vanish locally. In this example, however, the integrand 
goes to zero at spatial infinity in the extra dimensions 
faster than $1/r$ due to the asymptotic power law suppression of the 
metric (\ref{eqn:metric_AdS}). Therefore, using the Stokes's theorem, 
the domain wall tensions also vanish
\begin{eqnarray}
Z_m & = & 0. 
\label{eqn:Zm_eq_0}
\end{eqnarray}
Given $Y_m$ $=$ 0 in eq. (\ref{eqn:Ym_eq_0}), we find that 
all the topological contribution to the action vanish. 
Namely, the four-dimensional cosmological constant is zero. 
This is consistent with our ansatz of the
flat four-dimensional spacetime in eq. (\ref{eqn:metric-ansatz}).

The discussion here is based on a qualitative consideration. 
In particular the behavior of the solution inside the walls 
and the junction has not been obtained. 
We need to study the first order equations numerically 
to know the precise behavior of the solution. 

We normally need a fine tuning of parameters to obtain a vanishing
four-dimensional cosmological constant. This is also the case in our
model. The source of the fine tuning in our model 
lies in choosing the particular form of the scalar potential 
(\ref{eqn:potential}) rather than the detail of the function $W$. 
The same situation has been observed in five-dimensional models
with supergravity-motivated scalar potential in Refs. 
\cite{DeWolfe-etal,Verlinde}. At present, no symmetries to guarantee
the form of the potential (\ref{eqn:potential}) have been found. 
So we cannot solve the cosmological constant problem in our model.

\vspace{5mm}
%
%
%
\section{Gravity localization}
%

In this section we study the fluctuations of the metric defined by
\begin{eqnarray}
ds^2 & = & g^{(0)}_{MN} dx^M dx^N + h_{\mu \nu} dx^{\mu} dx^{\nu},
\label{eqn:h_munu}
\end{eqnarray}
where the first term in the right-hand side is the background metric
in eq. (\ref{eqn:metric-ansatz}). 
In this analysis, we focus on the transverse traceless modes 
which satisfy $h^{\mu}_{\ \mu}$ $=$ $\del^\mu h_{\mu \nu}$ $=$ 0.

In order to obtain the linealized equation for the fluctuation,
we expand the action (\ref{eqn:action}) 
in terms of $h_{\mu \nu}$ around the background 
metric up to the second order of the fluctuations. 
The second order terms are given by
\begin{eqnarray}
S^{(2)} & = & \int d^6 x \, \left[ -{\textstyle \frac{1}{8}} h^{\mu \nu}
\left\{ \Box_4  - \del_m^2 + 2 (\del_m^2 \sigma)
+ 4 (\del_m \sigma)^2 \right\} h_{\mu \nu} \right. \nonumber \\
 &   & + \left. \left\{ \del_m^2 \sigma
+{\textstyle \frac{3}{2}}(\del_m \sigma)^2 
+ {\textstyle\frac{1}{4}}( K_{ij^*} \del_m \phi^i \del_m \phi^{j*}
+ e^{2 \sigma} V ) \right\} h^{\mu \nu} h_{\mu \nu} \right],
\label{eqn:S_2}
\end{eqnarray}
where upper Lorentz indices are defined by the flat metric $\eta_{\mu \nu}$. 
The terms in the second curly bracket in the right-hand side cancel
because of the first order equations (\ref{eqn:BPS1})-(\ref{eqn:BPS3}). 
Variation of $S^{(2)}$ with respect to $h_{\mu \nu}$ gives rise to 
the linearized equation of motion for the metric fluctuation
\begin{eqnarray}
\left[\Box_4  - \del_m^2 + 2 (\del_m^2 \sigma)
+ 4 (\del_m \sigma)^2 \right] h_{\mu \nu}= 0.
\label{eqn:eq_fluctuation}
\end{eqnarray}
For the fluctuation with the four-dimensional dependence of a 
plane wave $e^{i p x}$, this equation can be written
as a two-dimensional Schr\"{o}dinger equation 
$(-\del_m^2 + V_{\rm QM})$ $h_{\mu \nu}$ $=$ $p^2 h_{\mu \nu}$ 
with a potential
\begin{eqnarray}
V_{\rm QM} & = & 2 (\del_m^2 \sigma) + 4 (\del_m \sigma)^2.
\label{eqn:Schrodinger_pot}
\end{eqnarray}
In the following we again consider the quartic function $W$ in 
eq. (\ref{eqn:superpot}) and the minimal K\"{a}hler potential. 
Using eq. (\ref{eqn:metric_AdS}), the asymptotic behavior of 
this potential far from the junction is given by
\begin{eqnarray}
V_{\rm QM} & \longrightarrow & \frac{6}{[{\rm Re}(\omega^{*n}z)]^2}.
\label{eqn:pot_asymptotic}
\end{eqnarray}
In five-dimensional models, the solutions to the corresponding
Schr\"{o}dinger equation for $p^2$ $>$ 0 are described by the Bessel
functions $\sqrt{x} J_2(px)$ and $\sqrt{x} Y_2(px)$ 
\cite{Randall-Sundrum-2,DeWolfe-etal}. In our case,
the equation involves the two variables, and the complete solution
is not available. However the potential
(\ref{eqn:pot_asymptotic}) shows that outside the wall, 
the Schr\"{o}dinger equation is essentially one dimensional equation. 
Therefore the solutions for $p^2$ $>$ 0 
can be described by the Bessel functions
$\sqrt{x} J_{5/2}(px)$ and $\sqrt{x} Y_{5/2}(px)$ 
outside the wall, where $x$ $=$ ${\rm Re}(\omega^{*n}z)$. 
These can be written by trigonometric functions only. 
The similar solutions have been found in Ref. \cite{string-like-defect}.

In the case of the massless fluctuation $p^2$ $=$ 0, the solution 
to eq. (\ref{eqn:eq_fluctuation}) is given by \cite{Carroll-etal}
\begin{eqnarray}
h_{\mu \nu} = e^{2\sigma} e^{i p x}\eta_{\mu \nu}. 
\label{eqn:massless_graviton}
\end{eqnarray}
This means that the massless fluctuation has the same configuration 
as the background metric in the extra dimensions.
From eq. (\ref{eqn:metric_AdS}) and our assumption on the
wall profile, we find that 
the massless graviton is localized on the junction. 
This agrees with the result in Ref. \cite{Carroll-etal}. 
We see that 
the massless mode (\ref{eqn:massless_graviton}) is normalizable
on our curved background
\begin{eqnarray}
\int dx^5 dx^6 \, e^{2 \sigma} h^{\mu \nu} h_{\mu \nu} & < & \infty. 
\label{eqn:normalizable}
\end{eqnarray}

In the transverse traceless components which we have considered,
there are no tachyonic modes. 
This is shown by writing eq. (\ref{eqn:eq_fluctuation}) as 
follows \cite{DeWolfe-etal}
\begin{eqnarray}
Q_m^{\dagger} Q_m h_{\mu \nu} = p^2 h_{\mu \nu},
\label{eqn:susy_QM}
\end{eqnarray}
where $Q_m$ $\equiv$ $-\del_m$ $+$ $2(\del_m \sigma)$. 
In the flat space, $Q_m^{\dagger}$ $\equiv$ $\del_m$ $+$ 
$2(\del_m \sigma)$ is the adjoint of $Q_m$. Similar equation
appears in supersymmetric quantum mechanics. 
The solutions to the Schr\"{o}dinger equation for $p^2$ $<$ 0 
are described by the modified Bessel functions 
$\sqrt{x} I_{5/2}(px)$ and $\sqrt{x} K_{5/2}(px)$. 
Following the discussion in Ref. \cite{DeWolfe-etal}, we can see that 
normalizable modes always satisfy $p^2$ $\geq$ 0 for the
transverse traceless components. 

\vspace{5mm}
%
%
\section{Summary}
%
%
In summary, we have studied a six-dimensional gravity coupled with 
complex scalar fields.
With a supergravity-motivated scalar potential, the domain wall 
junction solutions localize a four-dimensional massless graviton. 
We have shown that unlike the global supersymmetric model, 
contributions to the junction tension cancel locally with 
gravitational contributions. 
The wall tension vanishes due to the metric suppression.

%
%
\vspace{5mm}

%
%
%
\section*{Acknowledgements}
%
%

The author would like to thank L. Roszkowski and H.B. Kim for 
useful conversations. 
This work was supported in part by PPARC grant PPA/G/S/1998/00646.

\vspace{5mm}
%
%

%
%
%
%
\end{document}